\documentclass[reprint,twocolumn,aps,prB,amsmath,amssymb,floatfix,superscriptaddress,longbibliography]{revtex4-1}
\usepackage{graphicx}
\usepackage{epsfig}
\usepackage{bm}
\usepackage{dcolumn}
\usepackage{color}
\usepackage{physics}
\usepackage{float}
\usepackage{amsmath}
\usepackage[colorlinks,urlcolor=cyan,citecolor=blue,linkcolor=magenta]{hyperref}
\makeatletter
\newcommand*{\rom}[1]{\expandafter\@slowromancap\romannumeral #1@}
\makeatother
\usepackage{tikz}
\usepackage{subfigure}

\begin{document}

\preprint{APS/123-QED}
\preprint{This line only printed with preprint option}

\title{Emergent chiral toplogical point gaps in a non-Hermitian quasiperiodic Su-Schrieffer-Heeger model}

\author{Zhi-Bin Liang}
\affiliation {Key Laboratory of Atomic and Subatomic Structure and Quantum Control (Ministry of Education), Guangdong Basic Research Center of Excellence for Structure and Fundamental Interactions of Matter, School of Physics, South China Normal University, Guangzhou 510006, China}

\affiliation {Guangdong Provincial Key Laboratory of Quantum Engineering and Quantum Materials, Guangdong-Hong Kong Joint Laboratory of Quantum Matter, Frontier Research Institute for Physics, South China Normal University, Guangzhou 510006, China}

\author{Shan-Zhong Li}
\affiliation {Key Laboratory of Atomic and Subatomic Structure and Quantum Control (Ministry of Education), Guangdong Basic Research Center of Excellence for Structure and Fundamental Interactions of Matter, School of Physics, South China Normal University, Guangzhou 510006, China}

\affiliation {Guangdong Provincial Key Laboratory of Quantum Engineering and Quantum Materials, Guangdong-Hong Kong Joint Laboratory of Quantum Matter, Frontier Research Institute for Physics, South China Normal University, Guangzhou 510006, China}

\author{Runze Li}
\affiliation{Centre for Quantum Physics, Key Laboratory of Advanced Optoelectronic Quantum Architecture and Measurement (MOE), School of Physics, Beijing Institute of Technology, Beijing, 100081, China}

\affiliation{Beijing Key Lab of Nanophotonics \& Ultrafine Optoelectronic Systems, School of Physics, Beijing Institute of Technology, Beijing, 100081, China}

\author{Zhi Li}
\email[Corresponding author: ]{lizphys@m.scnu.edu.cn}
\affiliation {Key Laboratory of Atomic and Subatomic Structure and Quantum Control (Ministry of Education), Guangdong Basic Research Center of Excellence for Structure and Fundamental Interactions of Matter, School of Physics, South China Normal University, Guangzhou 510006, China}

\affiliation {Guangdong Provincial Key Laboratory of Quantum Engineering and Quantum Materials, Guangdong-Hong Kong Joint Laboratory of Quantum Matter, Frontier Research Institute for Physics, South China Normal University, Guangzhou 510006, China}

\date{\today}

\begin{abstract}
We study a quasiperiodic Su-Schrieffer-Heeger lattice with staggered on-site gain-loss. The results reveal that on-site staggered gain-loss can effectively induce non-Hermitian topological gap without introducing imaginary phase. Further analysis exhibits that five different processes of topological phase transitions, including topological re-entrant phenomena, with the gain-loss intensity increasing. In addition, through the analysis of inverse participation ratios, normalized participation ratios, winding number and other indicators, we find that topological phase transitions occur synchronically with localized phase transitions. Finally, by investigating the properties of dual space eigenfunction, we reveal that the non-Hermitian topological point gaps predicted in this paper are chiral point gaps, which have a pair of skin modes along opposite directions simultaneously.
\end{abstract}
\maketitle

\section{Introduction}
As a milestone in condensed matter physics, the discovery of topological matter has shaped the development of quantum simulation~\cite{OBoada2015,TOzawa2019,DWZhang2018,EAltman2021}. Topological states have been found in various artificial systems, such as ultracold atoms~\cite{NGoldman2016,NRCooper2019,KWintersperger2020}, optical waveguide arrays~\cite{JKang2022,SLKe2019}, superconducting quantum circuits~\cite{EFlurin2017,XTan2018,WCai2019,AYoussefi2022,XLi2024}, Rydberg atomic arrays~\cite{SdeLeseleuc2019}, etc. In Hermitian case, topological phases can be classified according to dimension and symmetry of the system~\cite{CKChiu2016,SRyu2010}. Topological invariants, as the indicator of topological phases, can be used to describe the robust edge states on open boundaries~\cite{JMZeuner2015,CPoli2015,PPeng2016,HXu2016,SWeimann2017,MPan2018,HZhou2018,SYYao2018}. However, real materials are often affected by the environment, in which case the behavior of the system should be described by the Lindblad master equation~\cite{GLindblad1976,DManzano2020}. In the past ten years, the research on non-Hermitian systems has attracted wide attention and made very important progress. The non-Hermitian Hamiltonian can be described by the master equation, therefore, the non-Hermitian Hamiltonian can effectively describe the open system~\cite{PMVisser1995,FRoccati2022}. Just recently, the discovery of non-traditional topological boundary states in experiments on non-Hermitian systems has aroused interest in extending topological band theory to open systems~\cite{NHatano1996,NHatano1997,PWBrouwer1997,NHatano1998,MSRudner2009,KEsaki2013,HSchomerus2013,CYuce2015,SLonghi2015,TELee2016,DLeykam2017,CYin2018,SYao2018a,HShen2018,VMMartinez2018,CYuce2018,KTakata2018,FKKunst2018,ZGong2018,SYao2018b,YXiong2018,KKawabata2018,AGhatak2019,DNakamura2024}. In non-Hermitian topological systems, the non-Hermitian skin effect~\cite{ZGong2018,SYao2018b} and the topological point gap~\cite{XJZhang2022,DNakamura2023,GHwang2023,KKawabata2019} have been found successively.

Quasiperiodic lattice, as a system between perfect periodic and completely random disorder model, has the long range correlation~\cite{PGHarper1955,SAubry1980}
. Aubry-Andr\'e (AA) model, as one of the most typical one-dimensional quasiperiodic lattices, is widely used to study Anderson transition~\cite{SAubry1980,JBSokoloff1984,GRoati2008,YLahini2009,JBiddle2010,XPLi2015,SGaneshan2015,TLiu2021,YZhang2022,QLin2022,BFZhu2023,SZLi2024b,GJLiu2024}. In Hermitian case, by a specific mapping, the AA model can be equivalent to a two-dimensional quantum Hall system in a square lattice, which presents important topological transport characteristics~\cite{YEKraus2012,LJLang2012,SGaneshan2013,MVerbin2013}. Recent studies have shown that a non-Hermitian topological point gap emerges in non-Hermitian system~\cite{SLonghi2019,YQZheng2025,YLiu2020,TLiu2020,YLiu2021,SLonghi2021,TLiu2022,XXia2022,QBZeng2020b,QBZeng2020a,XCai2022,XCai2021,SZLi2024a,LWang2024a,LWang2024b}. Previous studies have also revealed that the emergence of topological point gaps in quasiperiodic lattice is always accompanied by the localized phase, which means topological point gap corresponds to the skin effect in dual space~\cite{SLonghi2019,YQZheng2025}. 

Note that, in order to induce topological point gap phases in non-Hermitian quasiperiodic lattice, imaginary phases must be introduced~\cite{SLonghi2019}. A curious question naturally arises: Can non-Hermitian topological point gaps be induced by other means? Besides, in general, the localized wave function will remain localized after a Anderson transition. However, recent reports point out that certain constraints, such as staggered on-site potential~\cite{APadhan2022a}, interpolate on Fibonacci models~\cite{VGoblot2020}, non-diagonal modulation~\cite{DAMiranda2024}, etc., can be imposed on the quasiperiodic system to implement the re-entrant localization phenomenon. These new findings have gained extensive attention and brought about the second curious question to us: Whether the properties of the re-entrant localization can be used to construct systems with re-entrant non-Hermitian topological transitions?

\begin{figure}[htbp]
    \centering
    \includegraphics[width=8.5cm]{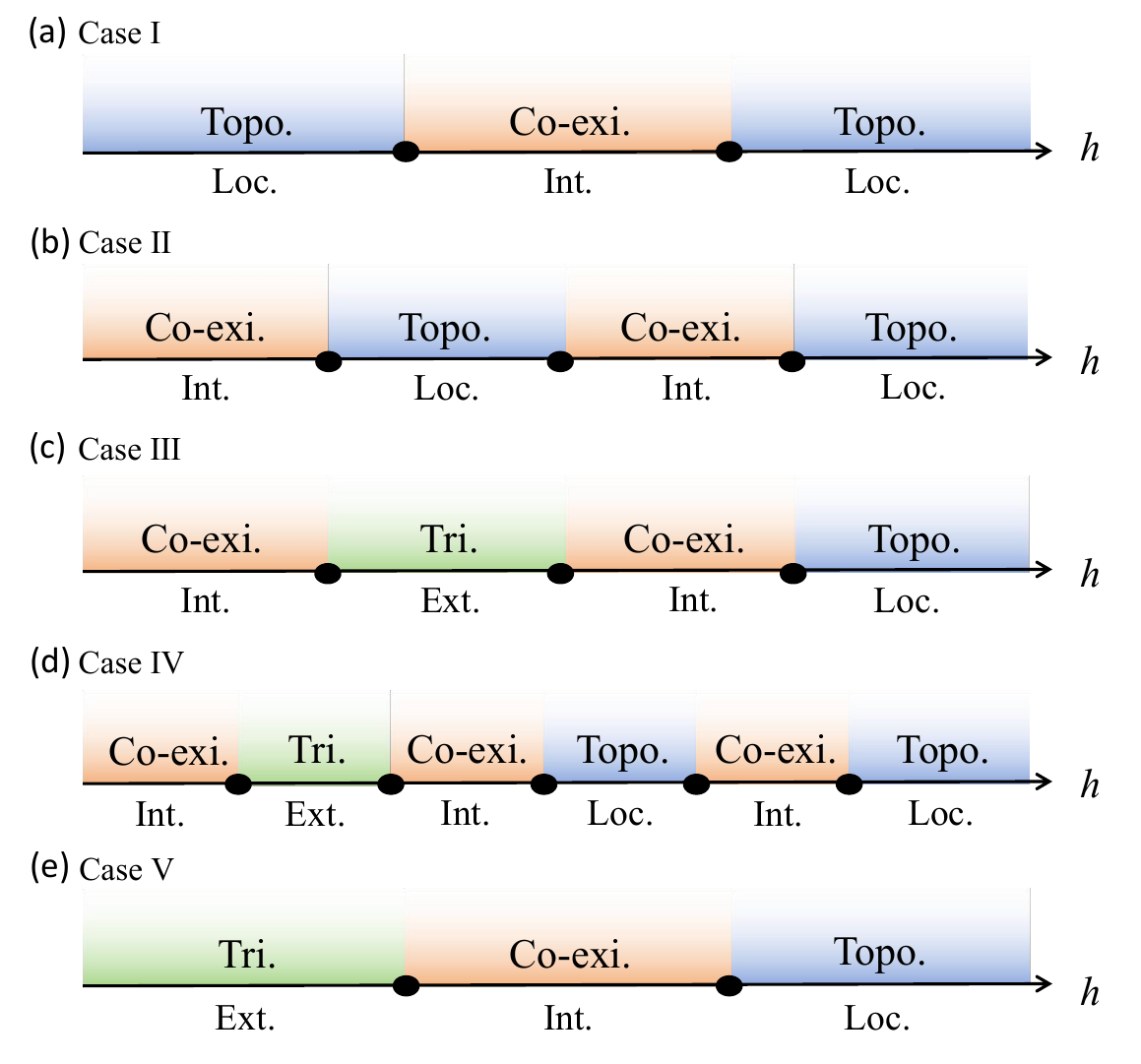}
    \caption{Five possible topological phase transition processes.} 
    \label{MF}
\end{figure}

The answers to the above two questions are both affirmative. The main finding of this paper is that the introduction of staggered gain-loss on-site energy can induce topological point gaps without imaginary phase, and the re-entrant phenomenon will also emerge. The five possible topological phase transition processes of the model are summarized in Fig.~\ref{MF}.

The rest of the paper is structured as follows. In Section II, we introduce the model and the main pointer quantities. Section III focuses mainly on the topological and localization properties of five phase transition processes. In Section IV, we reveal the chirality of the point gap topology through the properties of wave function in dual space. Section V is the summary of the finding results.

\section{Model and key observables}\label{Sec.2}
\begin{figure}[phtb]
    \centering
    \includegraphics[width=8.5cm]{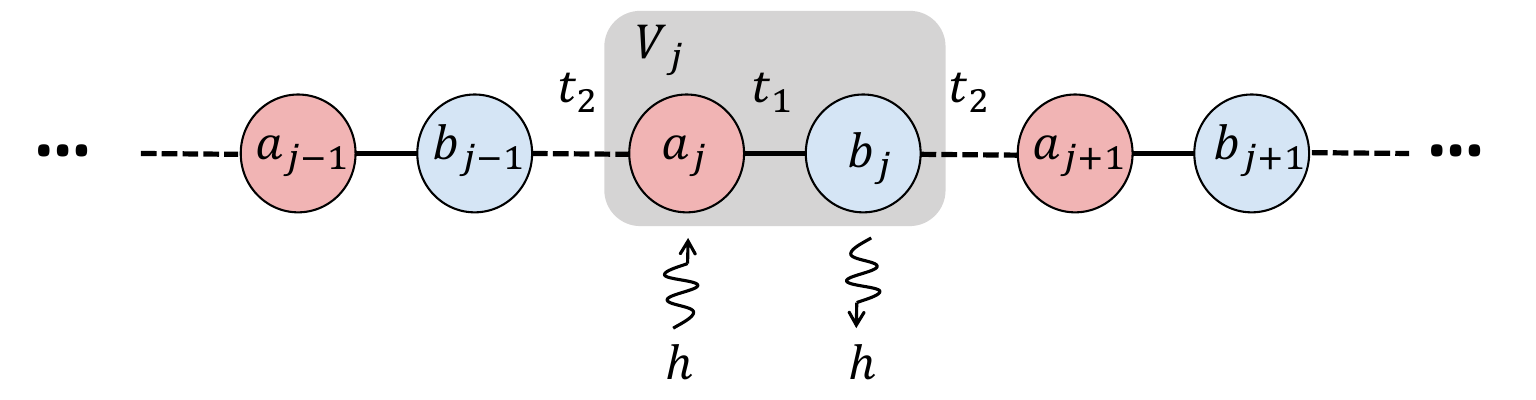}
    \caption{Schematic diagram of model~\eqref{Hamil}} 
    \label{F1}
\end{figure}

Let's start at a staggered gain-loss quasiperiodic Su-Schrieffer-Heeger model~\cite{WPSu1979}. The corresponding Hamiltonian reads
\begin{equation}\label{Hamil}
\begin{aligned}
H&=\sum_{j=1}^{L-1}t_{1}a_{j}^{\dagger}b_{j}+t_{2} a_{j+1}^{\dagger}b_{j}+H.c.+\sum_{j=1}^{L}V_{j}(a_{j}^{\dagger}a_{j}+b_{j}^{\dagger}b_{j})\\
&+\sum_{j=1}^{L}ih(a_{j}^{\dagger}a_{j}-b_{j}^{\dagger}b_{j})
\end{aligned}
\end{equation}
with
\begin{equation}
V_{j}=\lambda\cos(2\pi\alpha j+\theta),
\end{equation}
where $a_j^\dagger$ and $b_j^\dagger$ ($a_j$ and $b_j$) denote creation (annihilation) operators on $a,~b$ sublattices of the $j$-th site, respectively. $\lambda$ is the strength of the quasiperiodic potential. $t_1$ and $t_2$ correspond to the strength of intra and extra-hopping of unit cells, respectively (see Fig.~\ref{F1}). $h$ is the dissipation strength. For periodic boundary conditions (PBCs), the irrational number $\alpha$ can be selected as $\frac{F_{m}}{F_{m+1}}$ with $F_{m}$ being the $m$-th Fibonacci number. For open boundary conditions (OBCs), $\alpha=\lim_{m\rightarrow\infty}\frac{F_{m}}{F_{m+1}}=\frac{\sqrt{5}-1}{2}$. Without loss of generality, we set the global phase $\theta=0$. $L$ is the system size.\\ 

To study the topological phase transition, we calculate the winding number, i.e.,~\cite{SLonghi2019,HJiang2019}
\begin{equation}\label{omega}
    \omega(E_{b})=\lim_{L\rightarrow\infty}\frac{1}{2\pi i}\int_{0}^{2\pi} d\phi\frac{\partial}{\partial \phi} \ln\left[\mathrm{ det}(H-E_{b})\right],
\end{equation}
where $E_b$ is the base point. In other words, $\omega(E_{b})$ represents the number of times the spectrum winds a base energy $E_b$ when $\phi$ varies from 0 to $2\pi$. $\omega=0$ and $\omega=\pm1$ correspond to topological trivial and non-tivial phases, respectively. The topology of intermediate phase has two possibilities, i.e., the corresponding winding number can be equal to $0$ or $\pm1$.

Since the model~\eqref{Hamil} is a quasiperiodic lattice, Anderson transition will occur and we need to further analyze the corresponding localization characteristics. In concrete terms, we calculated the average inverse participation ratios (IPR) and normalized participation ratios (NPR), i.e.,
\begin{equation}
\begin{split}
    \overline{\xi}=\frac{1}{L} \sum_{\beta=1}^{L} \xi_{\beta},\\
    \overline{\zeta}=\frac{1}{L} \sum_{\beta=1}^{L} \zeta_{\beta},
\end{split}
\end{equation}
with
\begin{equation}
\begin{aligned}
    \xi_{\beta}&=\sum_{j=1}^{L}\sum_{s=a,b}\mid\psi_{j,s}(\beta)\mid^4,\\
    \zeta_{\beta}&=L(\sum_{j=1}^{L}\sum_{s=a,b}\mid\psi_{j,s}(\beta)\mid^4)^{-1},
\end{aligned}
\end{equation}
where $\psi_{j,s}(\beta)$ is the amplitude of the $\beta$-th eigenstate on sublattice $s$ of $j$-th unit cell. The extended state (localized state) corresponds to $\overline{\xi}=0 $ ($>0$) and $\overline{\zeta}>0 $ (=0). The intermediate phases are in between. 

In order to identify intermediate phases, one can define, based on the properties of average IPR $\overline{\xi}$ and NPR $\overline{\zeta}$, a physical quantity $\eta$ which takes the form~\cite{XLi2017,SRoy2021,APadhan2022b}
\begin{equation}
    \eta=\log_{10}[\overline{\xi}\times\overline{\zeta}].
\end{equation}
For the extended or localized phase, since the $\overline{\xi}$ or $\overline{\zeta}$ is proportional to $1/L$, $\eta<-\log_{10}L$. Here, we choose system size $L\backsim 10^3$ which means the system is being localized or extended when $\eta<-3$. In intermediate phase, the extended state and the localized state can coexist. That means both $\overline{\xi}$ and $\overline{\zeta}$ are finite and we get $-3<\eta<-1$.

    \begin{table}[htbp] 
    \renewcommand{\arraystretch}{2}
    \centering 
    \caption{Key indicators of states' localization and topology feature} 
    \label{T1} 
    \begin{tabular}{c c c c c} 
        \hline\hline
    State        ~ ~&~Topology      ~~&    ~~$\omega$         ~~&    ~~$\eta$        ~~&~~  $\Gamma$       \\ 
    \hline
    Loc.          ~~&~Topo.         ~~&    ~~$=\pm 1$         ~~&    ~~$<-3$         ~~&~~    $=0$        \\
    Ext.          ~~&~Tri.          ~~&    ~~$=0$             ~~&    ~~$<-3$         ~~&~~    $=1$           \\
    Int.          ~~&Co-exi.       ~~&    ~~$=0$~or~$\pm1$   ~~&    ~~$>-3$         ~~&~~  $=0$~or~$1$         \\
    \hline
    \end{tabular}
    \end{table}

Furthermore, the fractal dimension, as the core quantity reflecting the localization properties~\cite{YWang2020,YWang2022}, is defined as
\begin{equation}
    \Gamma =-\frac{\ln\overline{\xi} }{\ln L},
\end{equation}
$\Gamma=0$ ($\Gamma=1$) corresponds to a localized (extended) state. For convenience, we summarize the properties of key observables in Tab.~\ref{T1}.

\section{The topological and localization properties}\label{Sec.4}

\begin{figure}[phtb]
\centering
\includegraphics[width=8.5cm]{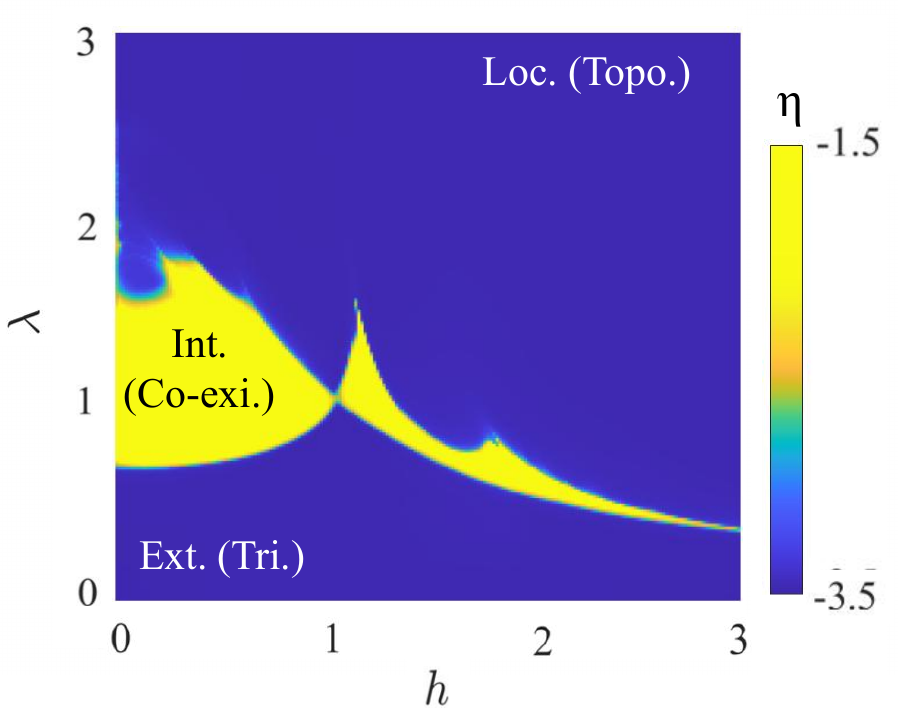}
\caption{$\eta$ phase diagram in $h$-$\lambda$ plane with $L=1220$.}
\label{F3}
\end{figure}

Based on the features of $\eta$, one can conduct an analysis on the system's localization properties. Specifically, the blue region ($\eta<-3$) corresponds to the localized or extended phase, while the yellow region corresponds to intermediate phase [see Fig.~\ref{F3}]. It can be seen clearly that the re-entrant localization phenomenon emerges. 

Generally, the localized phase transition and the non-Hermitian topological point gaps occur simultaneously. Upon careful examination of the phase diagram, one can find that five different re-entrant phase transitions occur one by one as the $\lambda$ increases, i.e., the region of $0<\lambda<0.75$, $0.75<\lambda<0.85$, $0.85<\lambda<1.05$, $1.05<\lambda<1.6$, $1.6<\lambda<1.8$. Below we discuss these five scenarios in detail. Without lose of generality, we set $t_1=t_2=1$ for convenience.

\subsection{Case I}
\begin{figure*}[htbp]
    \centering
    \includegraphics[width=14cm]{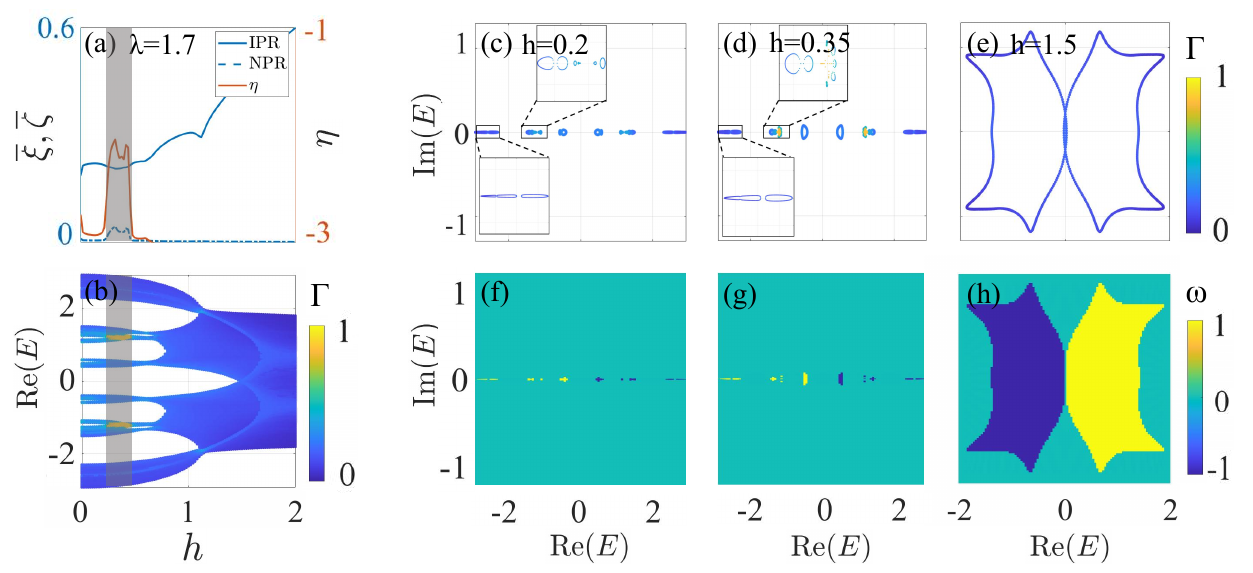}
    \caption{(a) IPR $\overline{\xi}$, NPR $\overline{\zeta}$ and $\eta$ versus $h$. (b) Fractal dimension $\Gamma$ versus $h$ as a function of real energies Re$(E)$. (c)-(e) The complex spectrum for different $h$. (f)-(h) The corresponding winding numbers. Throughout, we set $\lambda=1.7$, $L=1220$, and other parameters are marked.}
    \label{F4}
\end{figure*}

Firstly, we chose $\lambda=1.7$ as an example to discuss the first case. In Fig.~\ref{F4}, We show how average IPR $\overline{\xi}$(solid blue line), average NPR $\overline{\zeta}$(dashed blue line), and $\eta$(solid orange line) change with $h$. The results reveal that the system will gradually trasform from the localized phase into the intermediate phase, and finally enter the localized phase again with an inceasing $h$. The fractal dimension also confirms the emergence of reentrant localized phase transitions [see Fig.~\ref{F4}(b)]. Fig.~\ref{F4}(c-e) and Fig.~\ref{F4}(f-h) respectively exhibit the spectrum structure and winding number under the condition of $h=0.2,~0.35,~1.5$. In concrete terms, When $h=0.2$, the fractal dimension of all eigenstates tends to zero (blue dots), and looped non-Hermitian topological point gaps emerge. This is to say, the system has the characteristics of both localized phase and topological non-trivial phase under such condition. When $h=0.35$, a part of eigenstates' fractal dimension tends to zero (blue dots) and another part tends to 1(yellow dots). Meanwhile, there are two kinds of energy spectrum structures in the complex plane, i.e., the loop structure which corresponds to the eigenstates of non-Hermitian topological point gaps and the gapless spectrum which corresponds to the trvial states. All these results reveal that the system is in an intermediate state, i.e., there are both localized and extended states. From a topological point of view, the system appears as a coexistence phase of topological non-trival and topological trvial. Finally, under the condition of $h=1.5$, the area where the gap is closed opens again, so that only topological phases exist in the system again. The fractal dimension and winding number consistently confirm the emergence of this topological re-entrant phase transition.

\subsection{Case II}
\begin{figure*}[tbp]
    \centering
    \includegraphics[width=17cm]{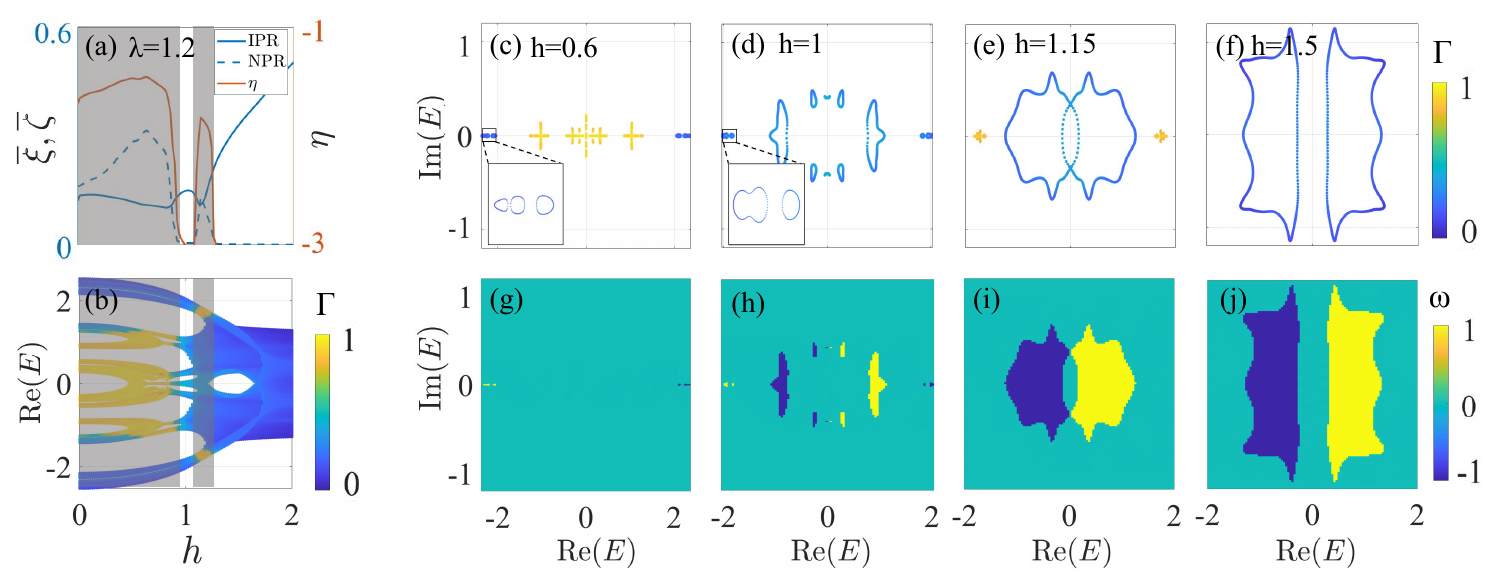}
    \caption{(a) IPR $\overline{\xi}$, NPR $\overline{\zeta}$ and $\eta$ versus $h$. (b) Fractal dimension $\Gamma$ versus $h$ as a function of real energies Re$(E)$. (c)-(f) The complex spectrum for different $h$. (g)-(j) The corresponding winding numbers. Throughout, we set $\lambda=1.2$, $L=1220$, and other parameters are marked.}
    \label{F5}
\end{figure*}

Next, let's turn to the second type of topological reentrant phase transition. For the second type of topologcial re-entrant phase transition, $\lambda=1.2$ is chosen as an example for further discussion. In Fig.~\ref{F5}(a) we draw $\overline{\xi}$, $\overline{\zeta}$, $\eta$ versus $h$. In Fig.~\ref{F5}(b), we plot the variation of the fractal dimension $\Gamma$ versus real energy Re$[E]$ and $h$. The gray area in the figure represents the topological Co-exi. phase. The results reveal that the system will start from the topological Co-exi. (Int.) phase, go through Topo. (Loc.) and Co-exi. (Int.) phases, and finally enter Topo. (Loc.) phase. We verify the emergence of the topological reentrant phase transitions by spectral structure, fractal dimension, and winding number [see Fig.~\ref{F5}(c-j)]. Note that, the same as in Fig.~\ref{F4}, topological point gaps (blue dots) and trivial spectrrum structures (yellow dots) appear simultaneously in the Co-exi. phases, which corresponds to localized and extended states, respectively. Therefore, from the perspective of localization characteristics, this is a intermediate phase.

\begin{figure*}[htbp]
    \centering
    \includegraphics[width=17cm]{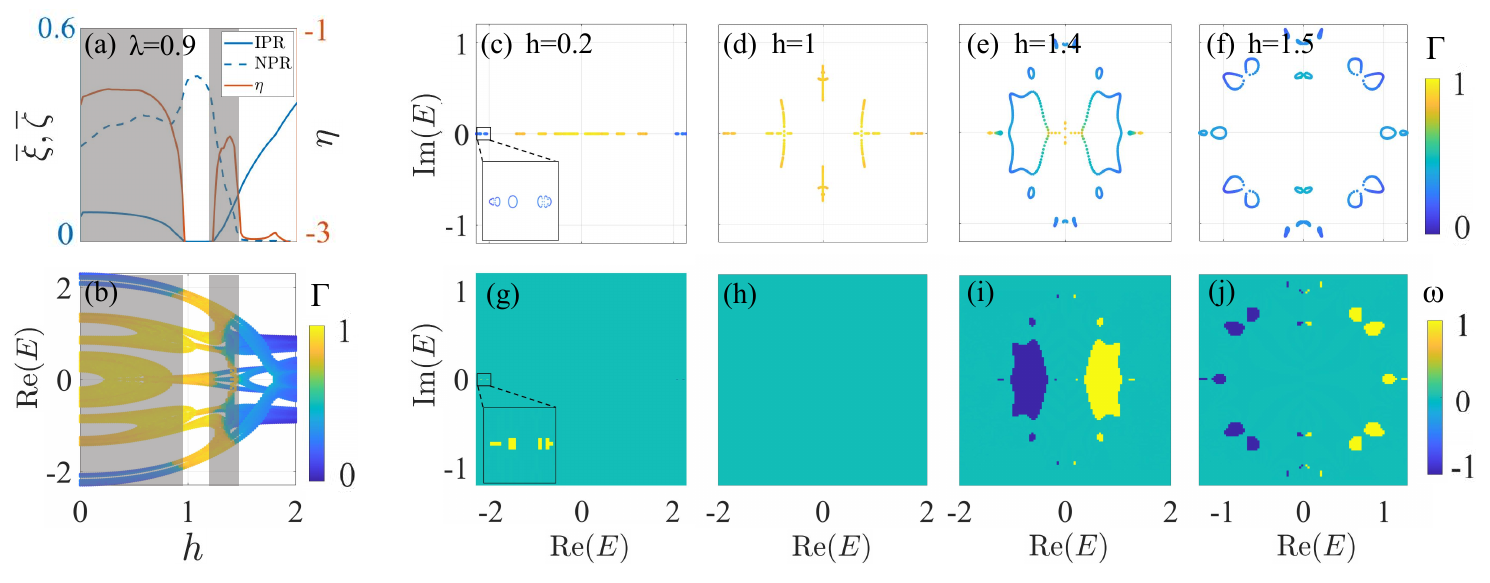}
    \caption{(a) IPR $\overline{\xi}$, NPR $\overline{\zeta}$ and $\eta$ versus $h$. (b) Fractal dimension $\Gamma$ versus $h$ as a function of real energies Re$(E)$. (c)-(f) The complex spectrum for different $h$. (g)-(j) The corresponding winding numbers. Throughout, we set $\lambda=0.9$, $L=1220$, and other parameters are marked.}
    \label{F6}
\end{figure*}
\subsection{Case III}

For the third type of topological reentrant phase transition, we choose the case of $\lambda=0.9$ as a typical example. Under such circumstances, we exhibit the properties of $\overline{\xi}$, $\overline{\zeta}$, $\eta$, and $\Gamma$ with respect to $h$. In concrete terms, when $h$ is small, the system will have both looped non-Hermitian topological point gaps and trivial gapless states [see Fig.~\ref{F6}(c) and (g)]. With the increase of $h$, the spectrum with non-Hermitian topological gap structures will gradually shrink and close, resulting in a pure tivial phase [see Fig.~\ref{F6}(d) and (h)]. If $h$ continues to increase, a part of eigenenergies will open the gap again, forming loop structures of non-Hermitian point gaps. Then, the system enters the topology Co-exi. phase again [see Fig.~\ref{F6}(e) and (i)]. Finally, the system will enter the pure topological phase with all eigenenergies being point gaps [see Fig.~\ref{F6}(f) and (j)]. Winding numbers reconfirm the emergence of this type of topological re-entrant phase transitions.

\subsection{Case IV}
\begin{figure}[htbp]
    \centering
    \includegraphics[width=8.5cm]{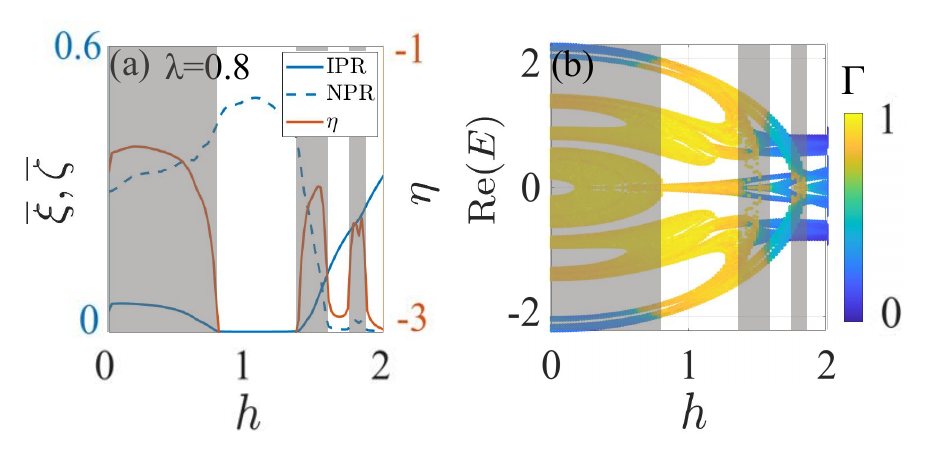}
    \caption{(a) IPR $\overline{\xi}$, NPR $\overline{\zeta}$ and $\eta$ versus $h$. (b) Fractal dimension $\Gamma$ versus $h$ as a function of real energies Re$(E)$.}
    \label{F7}
\end{figure}

Next, let's discuss the fourth type of topological reentrant phase transitions, which consist of six different phase regions [see Fig.~\ref{F7}]. Specifically speaking, as the parameter $h$ gradually increases, the system will go through phases of Co-exi., trival, Co-exi., Topo., Co-exi., Topo. phases one by one [see Fig.~\ref{MF}(d)]. From the perspective of localization features, the system experiences Int., Ext., Int., Loc., Int., and Loc. phases. By analyzing the behavior of average IPR $\overline{\xi}$, average NPR $\overline{\zeta}$, $\eta$ and the fractal dimension $\Gamma$, one can confirm the situation. In addition, since the system is below the yellow region of the overall phase diagram (see Fig.~\ref{F3}), the traditional extended state region is larger than the previous three typies of topological reentrant phase transitions.

\begin{figure*}[tbp]
    \centering
    \includegraphics[width=17cm]{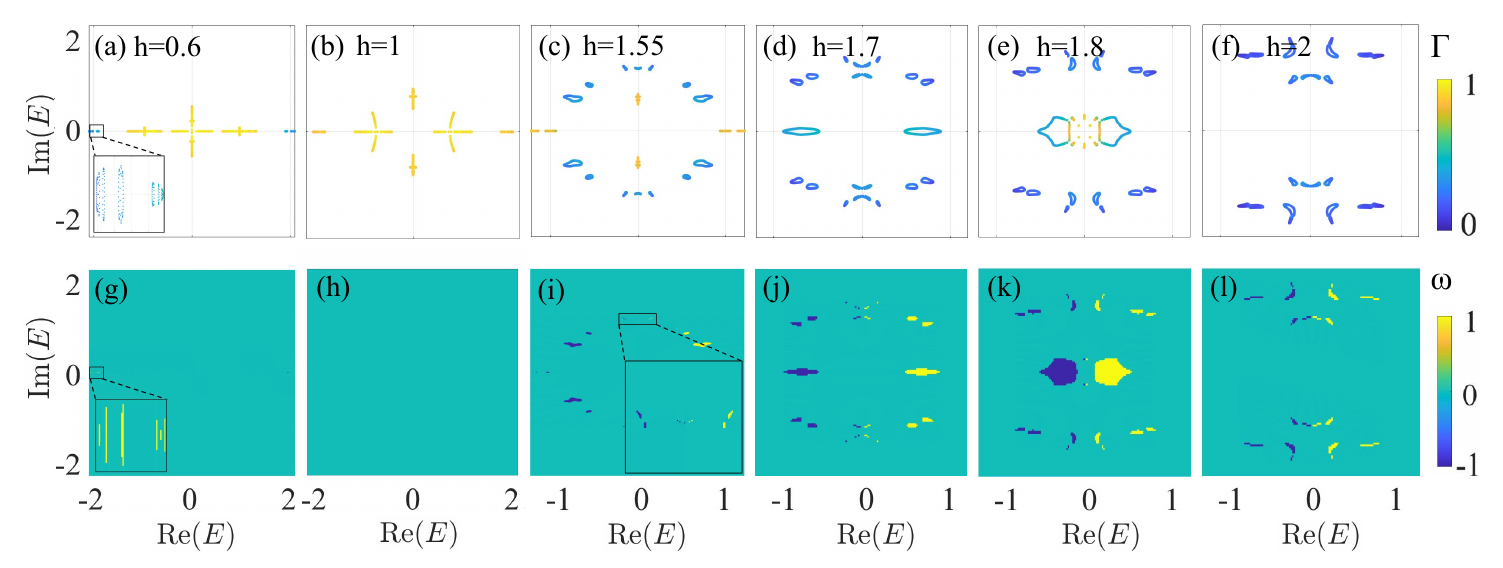}
    \caption{(a)-(f) The complex spectrum for different $h$. (g)-(l) The corresponding winding numbers. Throughout, we set $\lambda=0.8$, $L=1220$, and other parameters are marked.}
    \label{F8}
\end{figure*}

Furthermore, Fig.~\ref{F8} shows the eigenspectrum structure, fractal dimension and winding number corresponding to the system when parameter $h$ takes different values. By combining the results of the previous four types of topological reentrant phase transition, we find that the topological point gap structures will become more dispersed with the decrease of $\lambda$, and the size of loop structures in the complex plane will also become smaller [see Fig.~\ref{F8}(a)(c)(d)(e)(f)]. Because topological non-trivial spectrum become smaller, in Fig.~\ref{F8}(g)(i), we zoom in on the parts and see the non-zero winding numbers. 

\begin{figure*}[tbp]
    \centering
    \includegraphics[width=14cm]{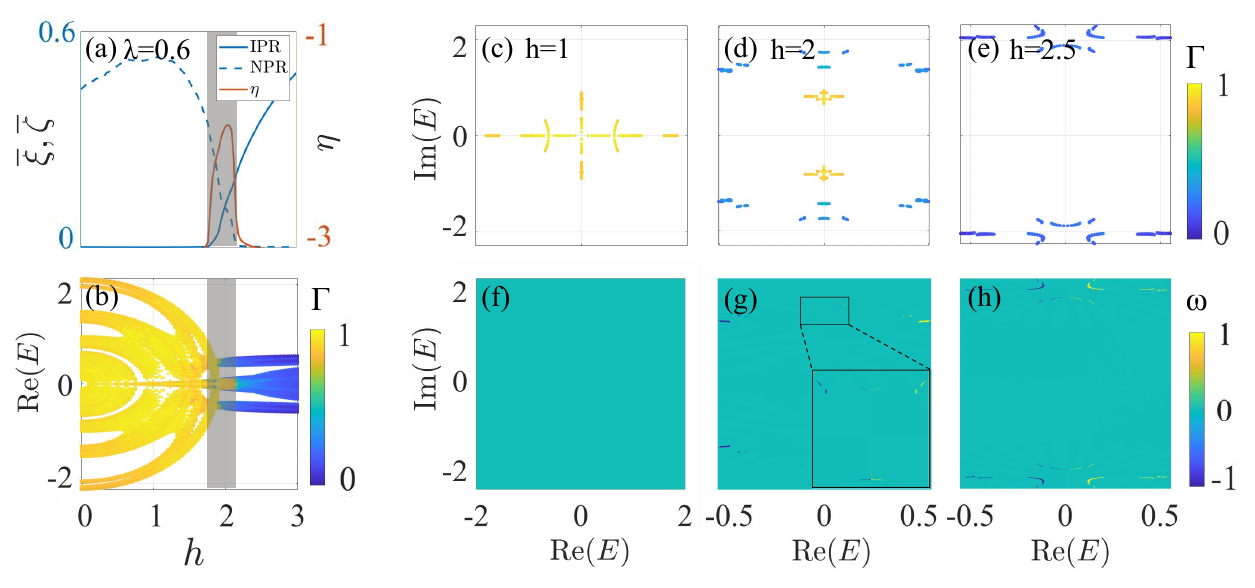}
    \caption{(a) IPR $\overline{\xi}$, NPR $\overline{\zeta}$ and $\eta$ versus $h$. (b) Fractal dimension $\Gamma$ versus $h$ as a function of real energies Re$(E)$. (c)-(e) The complex spectrum for different $h$. (f)-(h) The corresponding winding numbers. Throughout, we set $\lambda=0.6$, $L=1220$, and other parameters are marked.}
    \label{F9}
\end{figure*}

\subsection{Case V}
Finally, we show the fifth type of topological reentrant phase transition. Through similar analysis, one can find that the spectral structure, fractal dimension and winding number consistently confirm the phase transition from the trivial phase, through the Co-exi. phase, and finally into the topological phase in the system [see Fig.~\ref{F9}].

\section{Chiral Topology}\label{Sec.5}
\begin{figure}[htbp]
\centering
\includegraphics[width=8.5cm]{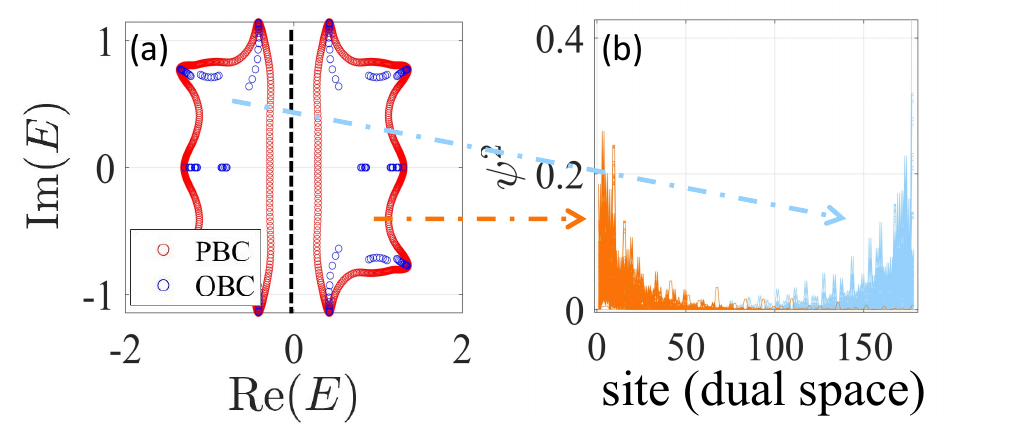}
\caption{(a) The eigenenergy in complex plane under OBCs (blue) and PBCs (red) with $\lambda=1.2$ and $h=1.5$. The system size $L=178$, $\alpha=\frac{55}{89}$ for OBCs and $L=1220$, $\alpha=\frac{377}{610}$ for PBCs. (b)The density distribution in dual space under OBCs. The blue (orange) line corresponds to left (right) skin modes.}
\label{F10}
\end{figure}

The emergence of topological point gaps in quasiperiodic systems are usually accompanied by the appearance of skin model in dual spaces. Now, we discuss the skin effect of dual space. The corresponding dual space Hamiltonian reads
\begin{equation}
\begin{aligned}
\label{Hami}
H_k &= \sum_{k=1}^{L} ( J + Je^{i2\pi\alpha k} ) a_{k}b_{k}^{\dagger} + ( J + Je^{-i2\pi\alpha k} ) a_{k}^{\dagger}b_{k}\\
&+ \sum_{k=1}^{L} ih ( a_{k}^{\dagger}a_{k} - b_{k}^{\dagger}b_{k} ) \\
&+ \sum_{k=1}^{L-1} \frac{\lambda}{2} ( a_{k+1}^{\dagger}a_{k} + b_{k+1}^{\dagger}b_{k} + \text{H.c.} ).
\end{aligned}
\end{equation}
We discuss the case of $\lambda=1.7,~h=1.5$, i.e., the case of Fig.~\ref{F4}(e). It is shown that when the left (right) skin mode emerge in dual space, the winding number $\omega=1$($=-1$) for the corresponding point gap. In addition, we also find that left and right skin modes simultaneously appear in the system, meaning that there are two skin modes in opposite directions at the same time [see Fig.~\ref{F10}(b)]. Next, we discuss the mechanism of dual space skin modes.

Previous studies have reported that the on-site dissipation induces a skin mode in non-Hermitian systems, which is known as the chiral skin effect~\cite{YYi2020}. To have a non-Hermitian skin effect, there are three feasible schemes:

I. Break the spatial inversion symmetry of a non-dissipative Hamiltonian;

II. Maintain the Hamiltonian's spatial inversion symmetry, and introduce the dissipation term which is antisymmetric with spatial inversion Hamiltonian;

III. Maintain the Hamiltonian's spatial inversion symmetry, and introduce the dissipation term which is commutative with the spatial inversion operator.

By analyzing the dual space Hamiltonian eq.~\eqref{Hami}, one can find that the mode proposed in this manuscript essentially conforms to the II scheme. In other words, the dual space skin effect caused by staggered gain-loss is indeed the chiral skin effect, so the corresponding non-Hermitian point gaps belong to the chiral topology class.

\section{Conclusion}\label{Sec.6}
In this paper, we propose a new method to induce the non-Hermitian topological point gaps, i.e., introducing staggered gain-loss. Different from the previous studies on non-Hermitian topology, this method does not need to introduce imaginary phase on the term of quasi-periodic potential, which will bring convenience to the experimental realization of the point gap. By analyzing the gain-loss intensity, we predict that five different topological reentrant phase transitions can emerge. Through numerical results of observable such as IPR, NPR, fractal dimension, winding number, etc., we confirm the existence of these five types of topologically re-entrant phase transitions. Furthermore, we prove that the nontrivial phases are chiral topologies, which correspond to a pair of chiral modes in opposite directions to the dual space. Since the induction method proposed in this paper is relatively easy to achieve experimentally (in the case of optical tweezers neutral atom arrays, only the auxiliary Rydberg laser is required to be interlaced), we hope that relevant phenomena can be observed on quantum simulation platforms such as optical tweezers atom arrays in the near future.

\section{acknowledgments}
We thank Yi-Qi Zheng and Jia-Ming Zhang for their insightful suggestions. This work was supported by the National Key Research and Development Program of China (Grant No.2022YFA1405300).

J.-M. Zhang, S.-Z. Li and R. Li contribute equally to
this work.

\end{document}